\documentclass[aps,prb,preprint,nopacs,superscriptaddress]{revtex4}

\usepackage{graphicx}
\usepackage{verbatim}
\usepackage{mathrsfs}
\usepackage{array}
\usepackage{epstopdf}
\newcolumntype{P}[1]{>{\centering\arraybackslash}p{#1}}

\usepackage{amsmath,amsfonts,amssymb}
\usepackage{graphicx}

\def\3{2.8in}    
\def\2{2.5in}
\def\4{3.0in}

\def \beq {\begin{equation}}
\def \eeq {\end{equation}}

\begin{document}

\title{Weyl Semimetals, Fermi Arcs and Chiral Anomalies (A Short Review)}

\author{Shuang~Jia}
\affiliation {International Center for Quantum Materials, School of Physics, Peking University, China}
\affiliation {Collaborative Innovation Center of Quantum Matter, Beijing,100871, China}
\author{Su-Yang Xu}\affiliation {Laboratory for Topological Quantum Matter and Spectroscopy (B7), Department of Physics, Princeton University, Princeton, New Jersey 08544, USA}
\author{M. Zahid Hasan}\affiliation {Laboratory for Topological Quantum Matter and Spectroscopy (B7), Department of Physics, Princeton University, Princeton, New Jersey 08544, USA}
\affiliation {Princeton Institute for Science and Technology of Materials (PRISM), Princeton University, Princeton, New Jersey 08544, USA}

\date{\today}

\maketitle
\textit{Physicists have discovered a novel topological semimetal, the Weyl semimetal, whose surface features a nonclosed Fermi surface while the low energy quasiparticles in the bulk emerge as Weyl fermions. Here they share a brief review of the development and present perspectives on the next step forward.}

\vspace{2cm}

Weyl semimetals are semimetals or metals whose quasiparticle excitation is the Weyl fermion, a particle that played a crucial role in quantum field theory but has not been observed as a fundamental particle in vacuum \cite{Weyl, hasanxubian, volovik, ciudad, herring_accidental_1937, Murakami2007, Wan2011, Yang, Burkov2011,HgCrSe, Weyl_review, TPT, TlBiSe, Huang2015, Weng2015, Hasan_TaAs, Hasan_FermiArc, TaAs_Ding,Chiral_anomaly_ChenGF, Chiral_anomaly_Jia, XuNbAs, ChenNatMat, Ding2, Belo}. Weyl fermions have definite chiralities, either left-handed or right handed. In a Weyl semimetal, the chirality can be understood as a topologically protected chiral charge. Weyl nodes of opposite chirality are separated in momentum space and are connected only through the crystal boundary by an exotic nonclosed surface state, the Fermi arcs. Remarkably, Weyl fermions are robust while carrying currents, giving rise to exceptionally high mobilities. Their spins are locked to their momentum directions due to their character of momentum space magnetic monopole configuration. The presence of parallel electrical and magnetic fields can break the apparent conservation of the chiral charge due to the chiral anomaly, making a Weyl metal, unlike ordinary nonmagnetic metals, more conductive with an increasing magnetic field. These new topological phenomena beyond topological insulators make new physics accessible and suggest potential applications, despite the early stage of the research
\cite{hasankane, hasanmoore, Ta3S2, ICSD, SrSi2, WT-Weyl, WMoTe-Weyl, MT-Weyl, MT-Weyl_2,weyl2_belo, LaAlGe, WT-ARPES-2, Ong_Chiral, ZrTe5_CA, ZrTe5_ARPES, ZrTe5_ARPES_2, Nonlocal, Arc_oscillation, Weyl_Floquet, TI_Floquet, Photocurrent, TNL}.

In this Commentary, we will review key experimental progress and present an outlook for future directions of the field. Through this article, we hope to expound our perspectives on the key results and the experimental approaches currently used to access the novel physics as well as their limitations. Moreover, while most of the current experiments are still focusing on the discovery of new Weyl materials and demonstration of novel Weyl physics such as the chiral anomaly, it is becoming clear that a crucial step forward is to develop schemes for achieving quantum controls of the novel Weyl physics by electrical and optical means. We discuss some theoretical proposals along these lines highlighting the experimental techniques and matching materials conditions that are necessary for realizing these research directions. 

\vspace{2cm}

\textbf{MATERIAL SEARCH}

\bigskip

Although the theory of Weyl semimetal has been around for a long time in various forms \cite{Weyl, herring_accidental_1937, Murakami2007, Wan2011}, its discovery had to wait until recent developments. This is because finding experimental realization requires appropriate materials simulation and characterizations. Historically, the first two material predictions, the pyrochlore iridates R$_2$Ir$_2$O$_7$ \cite{Wan2011} and the magnetically doped superlattice \cite{Burkov2011}, were both on time-reversal breaking (magnetic) materials. Perhaps influenced by the first works, for a long while, the community continued to focus on time-reversal breaking Weyl semimetals materials candidates \cite{HgCrSe, Weyl_review}. These candidates were extensively studied by many experimental groups. Unfortunately, these efforts were not successful in either demonstrating Fermi arcs or isolating the Weyl quasiparticles. Later on, it was realized that the ``time-reversal breaking'' route to finding suitable materials faces a number of obstacles such as the strong correlations in magnetic materials, the destruction of sample quality upon magnetic doping, and the issue of magnetic domains in photoemission experiments. On the other hand, for a long period of time, the inversion breaking Weyl materials were relatively unexplored. Following works on semimetallic state that locates at a topological phase transition point \cite{TPT}, Singh and collaborators \cite{TlBiSe} discussed this possibility, but they considered fine tuning of spin-orbit strength composition in TlBi(S$_{1-x}$Se$_{x}$)$_2$, which is challenging to fabricate. Looking for Weyl semimetal candidates in naturally occurring inversion breaking non-centrosymmetric single crystals can, however, avoid the difficulties described above. The advantages of this line of research for materials discovery were thoroughly discusses in reference \cite{Huang2015}, where the search was based on the Inorganic Crystal Structure Database of FIZ Karlsruhe \cite{ICSD}, which systematically records the lattice structure of crystals that have been synthesized over the course of a century. Following this approach and calculating the band structure of the materials that are likely semimetals, many experimentally feasible Weyl semimetal candidates have been identified \cite{Huang2015, Ta3S2, SrSi2, LaAlGe, Weyl_review}.

The theoretical prediction of Weyl semimetal states in the TaAs class of compounds were independently reported \cite{Huang2015, Weng2015}. The experimental realization of the first Weyl semimetal in TaAs followed soon after \cite{Hasan_TaAs},\cite{TaAs_Ding}. Both bulk Weyl cones and topological Fermi arc surface states were directly reported by photoemission providing a topological proof \cite{Hasan_TaAs} based on methods shown earlier \cite{Hasan_FermiArc}. The negative magneto-resistance, which serves as an early signature of the chiral anomaly, were reported by transport experiments in the TaAs family of materials \cite{Chiral_anomaly_Jia, Chiral_anomaly_ChenGF}. This family of materials was then experimentally shown to include NbAs and TaP \cite{Weyl_review, XuNbAs, Ding2, ChenNatMat, Belo}. A set of experimental criteria were developed for directly detecting topological Fermi arcs in Weyl semimetals without the need for a detailed comparison with band-structure calculation \cite{Belo}. To date, the TaAs family remains the only topological Weyl semimetals that have been clearly confirmed in experiments independent of comparison to bandstructure calculations \cite{Weyl_review, hasanxubian}.

The discovery of TaAs also established a feasible materials method to search for new Weyl semimetals that are more likely to be experimentally realizable. Shortly after TaAs, a number of new Weyl semimetal candidates along the same line of thinking were proposed. Primary examples include Ta$_3$S$_2$ \cite{Ta3S2}, SrSi$_2$ \cite{SrSi2}, Mo$_{1-x}$W$_x$Te$_2$ \cite{WT-Weyl, MT-Weyl_2, MT-Weyl, weyl2_belo, WMoTe-Weyl}, and LaAlGe \cite{LaAlGe}. In particular, SrSi$_2$ realizes a quadratic double Weyl state \cite{SrSi2} while Mo$_{1-x}$W$_x$Te$_2$ \cite{WT-Weyl, MT-Weyl_2, MT-Weyl, WMoTe-Weyl}, LaAlGe \cite{LaAlGe} and Ta$_3$S$_2$ \cite{Ta3S2} are the so-called type-II Weyl semimetals, where the Weyl fermion manifests as a strongly Lorentz violating tilted-over cone in the band structure \cite{WT-Weyl}. Experimental evidence of the bulk Weyl state has been shown in LaAlGe \cite{LaAlGe} and partially in Mo$_{1-x}$W$_x$Te$_2$ \cite{WT-ARPES-2, weyl2_belo} since in these systems Fermi arcs alone cannot distinguish between type-I and type-II states \cite{LaAlGe, weyl2_belo}.

\vspace{2cm}

\textbf{TRANSPORT RESPONSE and THE CHIRAL ANOMALY}

\bigskip

Weyl fermions are expected to exhibit various forms of quantum anomalies such as a topological Fermi arc and chiral anomaly. The chiral anomaly is important in understanding some structures of the standard model of particle physics which is based on the quantum field theory. A well-known case is the triangle anomaly associated with the decay of the neutral pion. Weyl semimetals provide an electronic route to realizing the chiral anomaly in condensed matter. Since the Weyl nodes are separated in momentum space, parallel magnetic and electric fields can pump electrons between Weyl nodes of opposite chirality that are separated in momentum space. This process violates the conservation of chiral charge and leads to an axial charge current, making a Weyl semimetal more conductive with an increasing magnetic field that is parallel to the electric field. 

\begin{figure*}
\centering
\includegraphics[width=16cm]{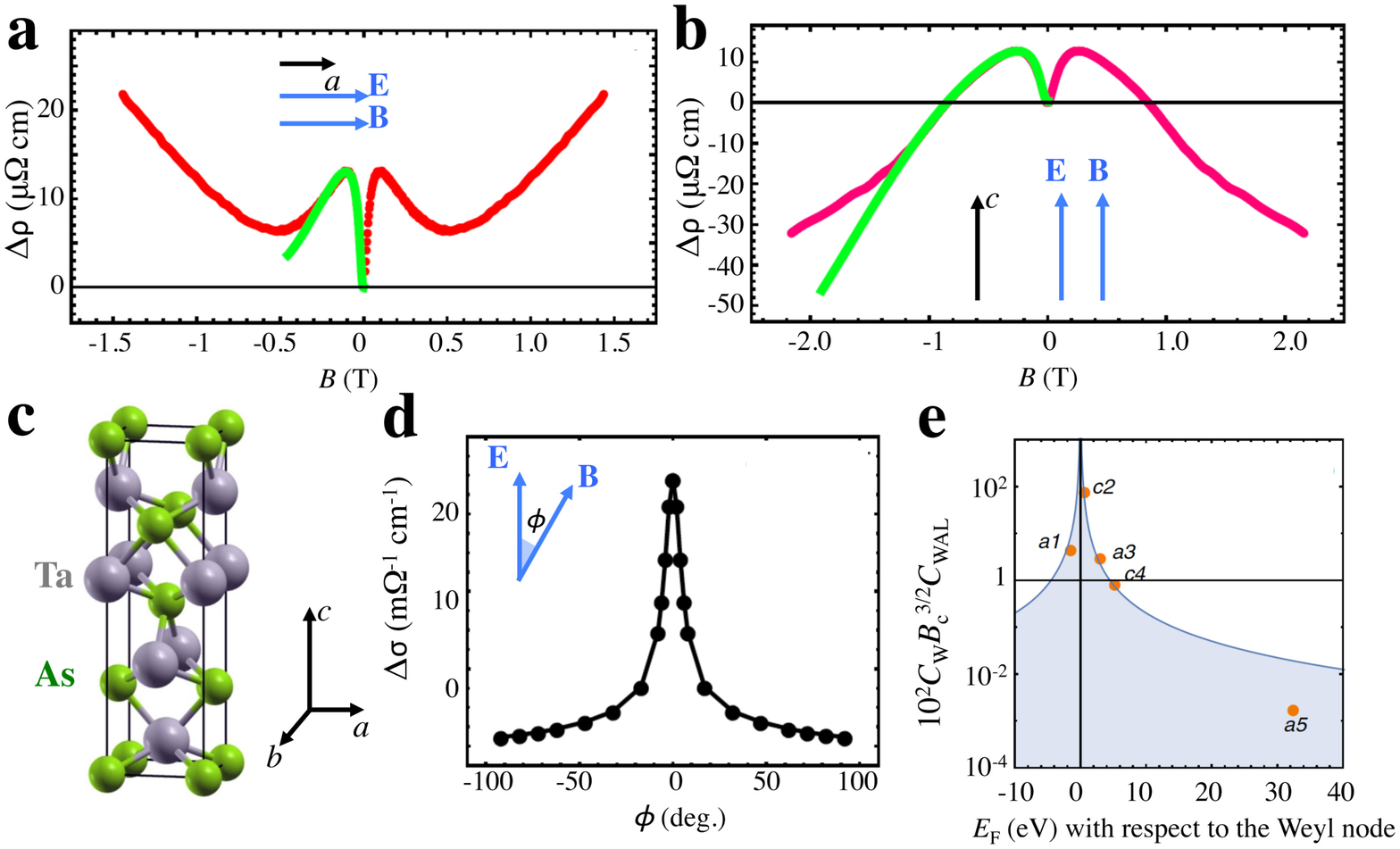}
\caption{\label{CA}\textbf{Signatures of Adler-Bell-Jackiw chiral anomaly in TaAs.} \textbf{a,b,}
Longitudinal magneto-resistance data of TaAs. The red lines are the data whereas the green lines are the theoretical fits. Background subtraction is discussed in Ref-\cite{Chiral_anomaly_Jia}.  \textbf{c,} The tetragonal crystal lattice of TaAs. \textbf{d,} Magnetoresistance data as a function of the angle between the $\vec{E}$ and $\vec{B}$ fields.  \textbf{e,} Dependence of the chiral coefficient $C_{\textrm{W}}$, normalized by other fitting coefficients, on chemical potential, $E_{\textrm{F}}$. Remarkably, the observed scaling behavior is $1/E_{\textrm{F}}^2$, as expected from the dependence of the Berry curvature on chemical potential in the simplest model of a Weyl semimetal, $\Omega \propto 1/E_{\textrm{F}}^2$. Adapted with modifications from Ref. \cite{Chiral_anomaly_Jia}.}
\end{figure*}

A clear identification of a Weyl semimetal with Fermi arcs in TaAs paved the way for the detection of the chiral anomaly. As shown in Figs.~\ref{CA}\textbf{a,b}, the magneto-resistance decreases with an increasing magnetic field in a finite range of fields. The observed negative longitudinal magneto-resistance serves as an important first signature of the chiral anomaly. In addition, a number of supporting evidence were reported in Ref. \cite{Chiral_anomaly_Jia}. These include: (1) A sharp dependence of the negative magneto-resistance on the angle between the electric and magnetic fields (Fig.~\ref{CA}\textbf{c}); (2) The presence of the negative magnetoresistance does not depend on the direction of the $\vec{E}$ field with respect to the crystalline axis; (3) The negative magneto-resistance shows a strong dependence on the chemical potential $E_{\textrm{F}}$. The chiral coefficient, which gives a measure of the magnitude of the chiral anomaly, diverges as the chemical potential approaches the energy of the Weyl node (Fig.~\ref{CA}\textbf{e}) consistent with its relation to the Berry curvature field. These transport results in combination with ARPES data modeled consistently together demonstrated existence of the chiral anomaly in TaAs driven by Weyl fermions. It is worth noting that chiral anomaly can also arise in a Dirac semimetal because the Dirac fermions can split into pairs of Weyl fermions of opposite chirality under an external magnetic field \cite{Ong_Chiral}. In most metals magneto-resistance is known to be positive; so a negative magneto-resistance, especially a longitudinal one, is quite rare. However, the chiral anomaly is not the only possible origin. It may arise through the giant magnetoresistance effect in magnetic materials. Also, with a high electronic mobility, it can occur through a purely classical geometric effect, the current-jetting. Furthermore, the negative magneto-resistance due to the chiral anomaly is a Berry curvature effect. However, a generic band structure without Weyl nodes can carry nonzero Berry curvature as long as time-reversal symmetry or inversion symmetry is broken. Whether such a generic band structure with nonzero Berry curvature yet without Weyl nodes can lead to a negative magneto-resistance is not theoretically understood. A recent experiment reported negative magneto-resistance in ZrTe$_5$ \cite{ZrTe5_CA}. There the interpretation was that ZrTe$_5$ is a Dirac semimetal and the Dirac fermion splits into pairs of Weyl fermions under the external magnetic effect. However, subsequent photoemission and STM experiments \cite{ZrTe5_ARPES, ZrTe5_ARPES_2} clearly revealed that the robust electronic groundstate of ZrTe$_5$ is that of a gapped semiconductor with a small ($\sim50$ meV) gap (not a Dirac semimetal). This suggests that a thorough theoretical understanding of the origins of the negative magneto-resistance in this material is lacking. Considering the complexity of the phenomenon, in order to achieve a convincing case, one needs to provide data that are sensitive to the unique properties of Weyl fermions and their systematic relation to the Berry curvature field. To date, the only dependence on the position of the chemical potential with respect to the energy of the Weyl node (see, Fig.~\ref{CA}\textbf{e}) that systematically tunes the Berry curvature field effectively is available in TaAs (Fig.~\ref{CA}\textbf{e}). Specifically, it showed that the magnitude of the chiral anomaly diverges over $\frac{1}{E_{\textrm{F}}^2}$. This demonstrates a Berry curvature monopole behavior, and thus constitutes a strong signature of the topological Weyl fermion physics.

\vspace{2cm}

\textbf{EMERGENT WEYL FERMIONS and TOPOLOGY OF FERMI ARCS}

\bigskip

\begin{figure*}
\centering
\includegraphics[width=15cm]{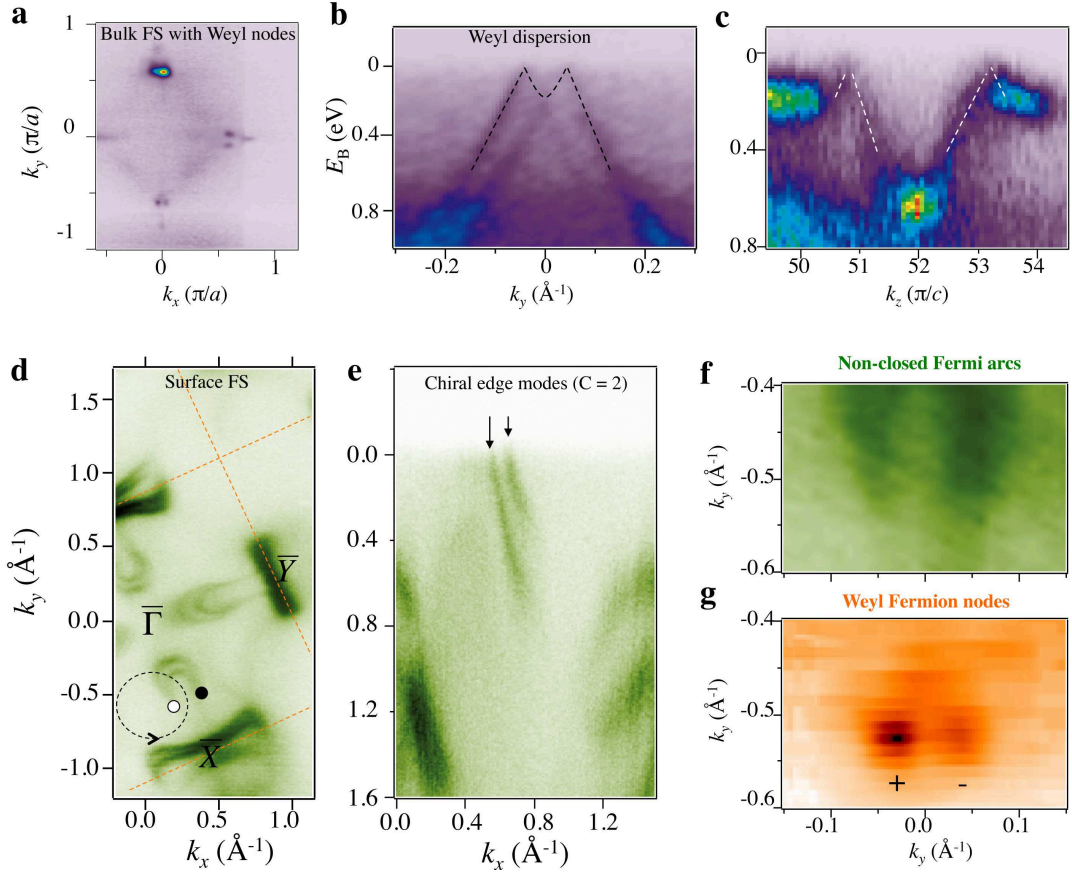}
\caption{\label{ARPES}\textbf{Observation of Weyl fermions and topological Fermi arcs.} \textbf{a,} ARPES measured $k_x,k_y$ bulk band Fermi surface projection of TaAs. The Fermi surface consists of discrete 0D points that arise from the Weyl nodes. \textbf{b,} In-plane energy dispersion ($E-k_{\|}$) that goes through a pair of Weyl nodes of opposite chirality. \textbf{c,} Out-of-plane energy dispersion ($E-k_{\perp}$) that goes through a pair of Weyl nodes. Panels (b and c) show that the bands disperse linearly away from the Weyl nodes both along the in-plane and the out-of-plane directions. \textbf{d,} Surface state Fermi surface of TaAs. We focus on the crescent shaped feature near the midpoint of the $\bar{\Gamma}-\bar{X}$ line. The Weyl semimetalic nature of TaAs can be demonstrated from the surface state band structure as measured in ARPES by drawing a loop in the surface Brillouin zone and counting the crossings to show a nonzero Chern number \cite{Hasan_TaAs}. The loop is shown by the dotted black line. \textbf{e,} The surface states along the loop: Two edge states of the same chirality seen in the momentum space cuts of the ARPES data reveal a Chern number of $2$ for this projection. ARPES methods for measuring Chern numbers are elaborated in Ref- \cite{Hasan_TaAs, Belo, weyl2_belo}. Note that for a conventional electron or hole pocket, such a counting argument should result in a null ($0$) value. Adapted from Ref. \cite{Hasan_TaAs}. }
\end{figure*}

Angle-resolved photoemission spectroscopy (ARPES) serves as a decisive method to experimentally demonstrate the Weyl semimetal state, since it can directly observe bulk Weyl fermions and surface Fermi arcs in the electronic band structure thus probing the bulk-boundary correspondence essential for proving a topological state of matter as demonstrated for topological insulators \cite{hasankane, hasanmoore, hasanxubian}. We review key ARPES observations that led to an unambiguous bulk-boundary topological demonstration of the Weyl semimetal state in TaAs. This demonstration methodology is a useful guideline for future research in this field as it can be more generally applied to discover Weyl or related topological states or novel topological quasipaticles in other materials. A Weyl node is a crossing point between the bulk conduction and valence band. Thus detection a Weyl node requires the ability to measure states along all three momentum space direction, $k_x$, $k_y$ and $k_z$. The $k_z$ resolution is achieved by tuning different incident photon energies. The existence of bulk Weyl cones and Weyl nodes in TaAs are shown by the following experimental observations in Figs.~\ref{ARPES}\textbf{a-c}. (1) The bulk conduction and valence bands touch at zero-dimensional (0D) isolated points (Fig.~\ref{ARPES}\textbf{a}). (2) These two bands disperse linearly away from the crossing point along all three momentum space directions ($k_x$, $k_y$, and $k_z$) (Figs.~\ref{ARPES}\textbf{b,c}). (3) The band crossings are not located at a high-symmetry point or along a high symmetry line but rather are observed at generic $k$ points of the Brillouin zone (Figs.~\ref{ARPES}\textbf{a-c}). The surface Fermi arcs are demonstrated by the following key observations in the experimental data. (4) The ARPES measured surface Fermi surface of TaAs (Fig.~\ref{ARPES}\textbf{d}) revealed a crescent-shaped feature, which consists of two curves that join each other at the two end points. (5) The surface band structure along a $k$ loop that encloses the end point of the crescent feature exhibits two chiral edge states, thus realizing a Chern number (C) of $+2$ (Fig.~\ref{ARPES}\textbf{e}) in analogy with the integer quantum Hall (IQH) chiral edge states. A Chern number is the band-structure invariant (momentum space wavefunction winding index) that describes the topology of the IQH state and gives a count of the protected edge states which is also proportional to the Hall conductivity. Compared to the IQH case with a single protected edge state (C=1) we now have 2 chiral edge states crossing Fermi level as seen in this momentum space cut in the ARPES data ((Fig.~\ref{ARPES}\textbf{e})), and therefore it is C=2 \cite{Hasan_TaAs, Belo, weyl2_belo}. This demonstrates that Weyl semimetals, in some aspects, feature momentum space properties that are higher dimensional analogs of quantum Hall-like states but without external magnetic field \cite{Weyl_review}. An intrinsic IQH state is not possible in 3D \cite{hasanxubian}. The observation of a nonzero Chern number in momentum space provides a clear and unambiguous demonstration of the topological origin of the Fermi arcs. The details of this Chern number method using ARPES is reported in Refs. \cite{Hasan_TaAs, weyl2_belo, TNL, Weyl_review}. (6) Finally, ARPES data showed that the terminations of the Fermi arcs coincide with the projections of the Weyl nodes, which demonstrated the bulk-boundary topological correspondence principle (Figs.~\ref{ARPES}\textbf{f,g}) and therefore the materials' topologically nontrivial nature.

A few remarks are in order: (1) The demonstration in TaAs only requires the above-described ARPES data along with established results in topological band theory. It does not rely on an agreement between the details of band dispersions and first-principles band structure calculations. Such an experimental demonstration of topological invariance is robust and reliable. (2) By contrast, a number of recent ARPES works have attempted to prove the existence of topological Fermi arcs solely based on the agreement between ARPES data and surface band structure calculations. While an agreement does provide supporting evidence, it cannot serve as a proof of the Fermi arcs. These issues are detailed in Refs. \cite{Hasan_TaAs, Weyl_review}. (3) In general, the bulk Weyl fermions or the surface Fermi arcs are two sets of independent but equally sufficient evidence for the Weyl semimetal state. Observing one of them with sufficient details, without the other, is sufficient to demonstrate a material as a type-I Weyl semimetal \cite{Belo}.

\vspace{2cm}

\textbf{QUANTUM CONTROL OF WEYL FERMIONS}
\bigskip

While ARPES and magneto-transport established the experimental foundation of the Weyl semimetal state, they are primarily aimed at demonstrating the topological physics including Weyl fermions, topological Fermi arcs, and the chiral anomaly. For this frontier, we note that all Weyl semimetals discovered so far belong to the inversion-breaking class. Thus finding a time-reversal breaking (most likely magnetic) Weyl semimetal is an important future task. It is also of interest to find simpler materials with fewer Weyl points. Material searches are under way to find the ``hydrogen atom'' versions of Weyl semimetals with the minimum number of Weyl points possible, either four Weyl nodes in inversion-breaking Weyl semimetals or two Weyl nodes in time-reversal breaking Weyl semimetals and topological nodal-line fermion materials such as PbTaSe$_2$ \cite{TNL}.

\begin{figure*}
\centering
\includegraphics[width=15cm]{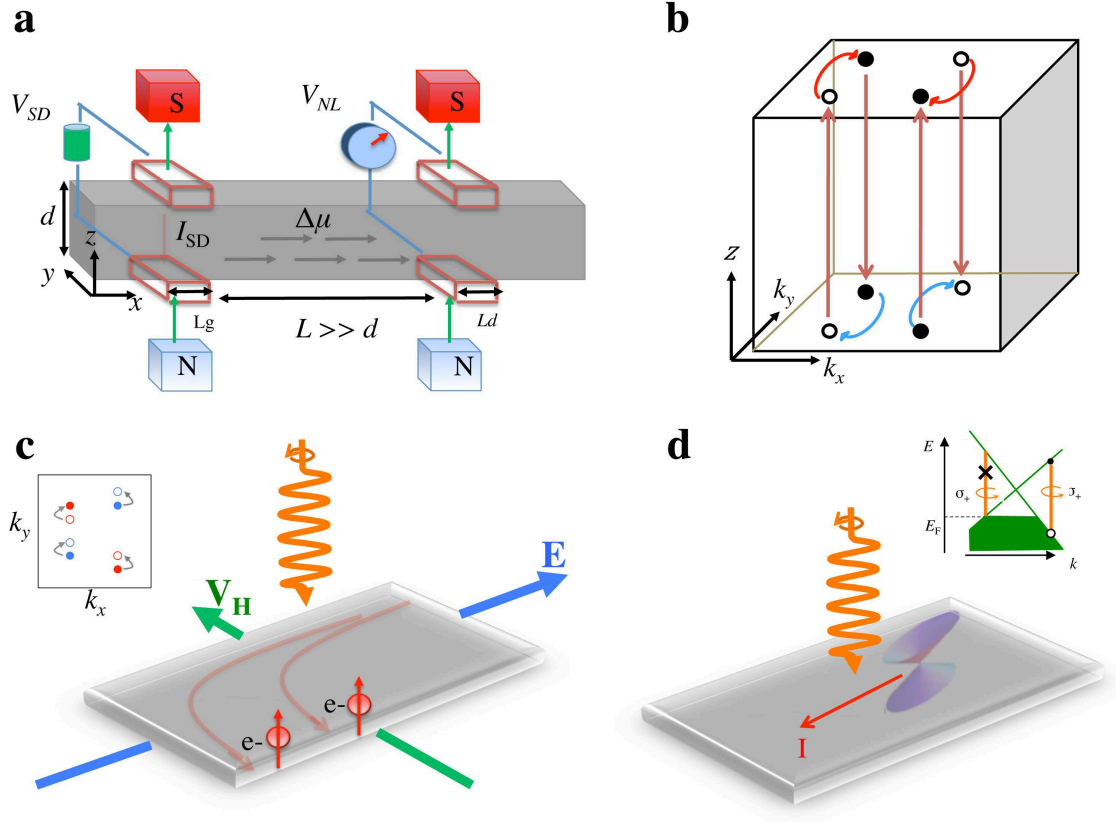}
\caption{\label{Device}\textbf{Electronic and optical control of Weyl fermions.} \textbf{a,}  Schematics of a nonlocal electrical transport device that utilizes the axial current arising from the chiral anomaly. \textbf{b,} On the surface of a Weyl semimetal, electrons exhibit an unusual path in real and momentum ($z-k_x-k_y$) space under an external magnetic field along the $z$ direction. Not all paths are shown in the figure. \textbf{c,} An intense circularly polarized light can break time-reversal symmetry and leads to a nonzero Hall conductivity via the Floquet effect in the absence of an external magnetic field. Inset: the solid (open) circles show the Weyl nodes in the absence (presence) of the light. \textbf{d,} A circularly polarized light can generate a large photogalvanic current in a Weyl semimetal that breaks inversion symmetry and any mirror symmetry. Inset shows the optical selection rule of right-handed circularly polarized light from the lower band to the upper band of the Weyl cone. Panels (a-d) are based on Refs. \cite{Nonlocal, Arc_oscillation, Weyl_Floquet, Photocurrent}. }
\end{figure*}

Here, we highlight the potential application of Weyl semimetals in device applications. Moving forward it is important to achieve systematic control over Weyl physics via electric and optical means so that these novel phenomena can potentially be harnessed in device settings. A recent theory work \cite{Nonlocal} proposed an electrical device which can utilize the dissipationless axial current due to the chiral anomaly. The chiral charge pumping effect (Fig.~\ref{CA}\textbf{f}) leads to a current, the axial current, which is believed to be nearly dissipationless. Any relaxation of the axial current requires scattering from one Weyl node to the other, which occupies a limited volume of the parameter space as it involves a large specific momentum transfer vector that connects the two Weyl nodes. As shown in Fig.~\ref{Device}\textbf{a}, in this device, it is proposed to have a thin and long ribbon of a Weyl semimetal sample. On the left, parallel magnetic and electric fields lead to the axial current, which can drift to the right and be detected in the form of a spontaneous voltage/current without applying an electric field. Another theory work \cite {Arc_oscillation} predicts a novel type of quantum oscillation that arises from the topological Fermi arc surface states. As shown in Fig.~\ref{Device}\textbf{b}, a surface electron occupying a state on the Fermi arcs has to travel back and forth across the bulk of the sample in the presence of an out-of-plane magnetic field. A direct consequence is that the period of the oscillation as a function of inverse magnetic field $\frac{1}{B}$ now depends on the thickness of the sample. The proposed experiments not only provide transport demonstration of the chiral anomaly and the surface arcs independent of ARPES or the negative magneto-resistance, but also realize novel device configurations that make use of the axial current and the Fermi arcs.

There have been a number of noteworthy theoretical proposals for achieving optical control of the novel Weyl physics. For example, Ref. \cite{Weyl_Floquet} proposed an optically induced anomalous Hall current in an inversion breaking Weyl semimetal. The authors showed that an intense circularly polarized light can break time-reversal symmetry, which shifts the $k$ space location of the Weyl nodes through the Floquet effect \cite{TI_Floquet}, and therefore leads to a nonzero Hall conductivity in the absence of an external magnetic field. (Fig.~\ref{Device}\textbf{c}). Another work \cite{Photocurrent} predicted a large photogalvanic current (as large as $10^7$ $\textrm{Am}^{-2}$ in TaAs \cite{Photocurrent}) in inversion-breaking Weyl semimetals. The proposed photocurrent is due to the optical selection rules (Fig.~\ref{Device}\textbf{d}) and it is different from the other proposal as it is a current induced only by shining light onto the sample without additional external electric (or magnetic) field. The large photocurrent can be utilized for fabricating high-sensitivity and broad bandwidth infrared detectors and for light harvesting schemes in solar cells. While these are only guidelines for potential applications, the plethora of unusual physics harbored by Weyl materials may lead to something far more exotic we have not thought of yet.

\textbf{Acknowledgements} Authors wish to thank I. Belopolski, S.-M. Huang, G. Bian, N. Alidoust, M. Neupane for comments and D. Haldane, I. Klebanov and E. Witten for discussion as a part of Princeton Summer School on New Insights Into Quantum Matter as a part of Prospects in Theoretical Physics Program at IAS. S.J. is supported by National Basic Research Program of China (Grant No.2014CB239302 and 2013CB921901). Works of S.-Y.X and M.Z.H. are supported by U.S. Department of Energy under Basic Energy Sciences (Grant No. DOE/BES DE-FG-02-05ER46200 and No. DE-AC02-05CH11231 at Advanced Light Source at LBNL) and Princeton University funds. M.Z.H. acknowledges visiting scientist user support from Lawrence Berkeley National Laboratory, PRISM and partial support from Moore foundation (GBMF4547).

\textbf{Additional information}
Reprints and permissions information is available online at www.nature.com/reprints.
Correspondence and requests for materials should be addressed to S.J. (gwljiashuang@pku.edu.cn), S.Y.X. (suyangxu@Princeton.edu) or M.Z.H. (mzhasan@princeton.edu).

\textbf{Competing financial interests}
The authors declare no competing financial interests.

\clearpage

\clearpage

\end{document}